\documentclass[journal,12pt,onecolumn]{IEEEtran}
\usepackage{cite,url,enumerate,graphicx}
\usepackage[cmex10]{amsmath}
\usepackage{amssymb,amsthm}

 \newcommand\bZ{\mathbb Z} 
\newcommand{\bbsm}{( \begin{smallmatrix}}      \newcommand{\besm}{\end{smallmatrix} )}
\newcommand{\bbm}{\begin{pmatrix}}      \newcommand{\bem}{\end{pmatrix}}
\newcommand\beq{\begin{equation}}  \newcommand\eeq{\end{equation}}
\newcommand\beqs{\begin{equation*}}  \newcommand\eeqs{\end{equation*}}
\newcommand\bep{\begin{IEEEproof}}  \newcommand\eep{\end{IEEEproof}}

\newcommand\mF{\mathcal F}  \newcommand\mC{\mathcal C} 
 \newcommand\mL{\mathcal L}

\newtheorem{theorem}{Theorem}%[section]
\newtheorem{lemma}{Lemma}
\newtheorem{proposition}{Proposition}
\newtheorem{corollary}{Corollary}
\newtheorem{conjecture}{Conjecture}

\newtheorem*{conjecture*}{Conjecture}
\newtheorem*{theorem*}{Theorem}

\begin{document}
\title{An improvement of the  asymptotic Elias bound for non-binary codes.}
\author{Krishna~Kaipa
        % <-this % stops a space
\thanks{K. Kaipa is with the Department of Mathematics at the  Indian Institute of Science and Education Research, Pune  411008, India (email: kaipa@iiserpune.ac.in). The author
was supported by the Science and Engineering Research Board, Govt. of India,
under Project EMR/2016/005578.}}
\maketitle
\begin{abstract}
For non-binary codes the  Elias bound is a good upper bound for the asymptotic information rate at low relative minimum distance, where as 
the Plotkin bound is better  at  high relative minimum distance. In this work, we  obtain a hybrid  of these bounds which improves both. 
This in turn  is based on the  anticode bound which is a hybrid of the Hamming and Singleton bounds and  improves both bounds.
 The  question  of convexity of the asymptotic rate function is an  important open question.  We conjecture a much weaker form of the convexity, and  we show that our bounds  follow immediately if we assume the conjecture.

\end{abstract}
\begin{IEEEkeywords}  information rate, size of a code,  anticode.
\end{IEEEkeywords}

\IEEEpeerreviewmaketitle

\section{Introduction} \label{i}
%\IEEEPARstart{L}{et} 

Let $A_q(n,d;\mL)$ denote  the maximum size of a code of length $n$, minimum distance at least $d$, and contained in  a subset $\mL \subset \mF^n$,   where 
  $\mF$ is an alphabet of finite size $q$. A central problem in coding theory is to obtain good upper and lower bounds for $A_q(n,d)=A_q(n,d;\mF^n)$.
 The asymptotic version of this quantity is the asymptotic information rate function: 
\beq \label{eq:alpha} \alpha(x) = \limsup_{n \to \infty} n^{-1} \log_q A_q(n, x n), \; x \in [0,1].\eeq
 The quantities $A_q(n,d;\mL)$ and $A_q(n,d)$ are related  by the inequality
 \beq  \label{eq:BE} A_q(n,d) \leq q^n A_q(n,d;\mL)/ |\mL|, \eeq
known as the Bassalygo-Elias lemma.
Taking $\mL$ to be a Hamming ball of diameter $w$, and choosing $w$ optimally  gives, at the asymptotic level the Hamming and the Elias  upper bounds. 
 \beq \label{eq:sp} \alpha_{H}(x) = 1 - H_q(x/2),    \; x \in [0,1]. \eeq
\beq \label{eq:E} \alpha_E(x)= \alpha_{H}(2 \theta(1 - \sqrt{1 - x/\theta} )),    \; x \in [0, \theta]. \eeq
The bound $\alpha_E$ is better than $\alpha_H$  for all $x$. Here $H_q(x)$ is the entropy function \eqref{eq:H},
and $\theta := 1 - q^{-1}$.

An anticode of diameter $w$ in $\mF^n$ is any subset of $\mF^n$ with Hamming diameter $w$. Let $A^*_q(n,w)$ denote the maximum size of an anticode of diameter at most $w$ in $\mF^n$.   In  contrast to the situation with $A_q(n,d)$, the quantity $A^*_q(n,d)$ was explicitly determined by  Ahlswede and Khachatrian in \cite{AK}. From their result, it is easy to determine the asymptotic quantity $ \alpha^*(x) = \lim_{ n \to \infty} n^{-1} \log_q A_q^*(n,xn)$. 
We actually do not need the results of \cite{AK}, however it is the main inspiration for this work.
Taking $\mL$ to be an $A^*_q(n,w)$ anticode in \eqref{eq:BE}, and choosing $w$ optimally, we get the following two bounds which improve $\alpha_H$ and $\alpha_E$ respectively.
\begin{theorem} \label{HSthm} (hybrid Hamming-Singleton  bound) 
 \[ \alpha_{HS}(x) =  \begin{cases}
1-  H_q(\tfrac{x}{2})\!\! &\text{if  $x \in [0,2/q]$} \\
 (1-x) H_q(1)\!\!  &\text{if  $ x \in [2/q,1]$} . \end{cases}\]
The bound $\alpha_{HS}$ improves the Hamming and the Singleton bounds. It is $\cup$-convex and continuously differentiable.
 \end{theorem}
 
 \begin{theorem}  \label{EPthm} (hybrid Elias-Plotkin bound)  Let $q >2$.
 \[  \alpha_{EP}(x) = 
  \begin{cases} 1 -H_q(\theta  - \sqrt{\theta^2 - x \theta})\!\!\!     &\text{if  $x \in [0,\tfrac{2q-3}{q(q-1)}]$}\\ 
 (\theta - x) \tfrac{ (q-1) H_q(1)}{q-2}\!\!\!  &\text{if $x \in[ \tfrac{2q-3}{q(q-1)},\theta]$} 
 \end{cases}   \]
 The bound $\alpha_{EP}$  improves  the Elias and Plotkin bounds. It is $\cup$-convex and continuously differentiable.
 \end{theorem}
It is not known if the function $\alpha(x)$ itself is $\cup$-convex, although  it  is tempting to believe that it is. We propose a weaker conjecture:
\begin{conjecture} \label{conj1}
The function $\tfrac{\alpha(x)}{\theta-x}$ is decreasing. In other words
\[ \alpha(t x + (1-t) \theta) \leq t \alpha(x) + (1-t) \alpha(\theta),    \; t \in [0,1].\] 
\end{conjecture}
 As evidence for this conjecture, we will show that  theorems \ref{HSthm} and \ref{EPthm} follow very easily if we admit the truth of the conjecture.
 
 The bound $\alpha_{EP}$ in Theorem \ref{EPthm} is an elementary and explicit correction to the classical Elias bound. 
 It does not however improve the upper-bounds obtained by the linear programming approach, like the second MRRW bound $\alpha_{MRRW2}$ (due to Aaltonen \cite{Aalt1}) or the further improvement of $\alpha_{MRRW2}$  due to Ben-Haim and Litsyn \cite[Theorem 7]{BL}.  The reasons for this are as follows: For small $\delta$ we have $\alpha_{EP}(\delta) = \alpha_E(\delta) \geq \alpha_{MRRW2}(\delta)$.  For large $\delta$,  the inequality $\alpha_{EP}(\delta) > \alpha_{MRRW2}(\delta)$ follows from the fact that $\alpha_{EP}(\delta)$ has a non-zero slope at $\delta=1-1/q$ where as the actual function $\alpha(\delta)$ and the bound $\alpha_{MRRW2}$ have zero slope at $\delta=1-1/q$. \\
 The paper is organized as follows. In section \ref{ac}, we collect some results on size of anticodes, which we use in section \ref{sec3} to prove Theorems \ref{HSthm} and \ref{EPthm}. We discuss Conjecture \ref{conj1} in section \ref{convx}.
\section{Size of anticodes} \label{ac}
We recall that $A^*_q(n,d)$ is the maximum size of an anticode of diameter at most $d$ in $\mF^n$. If we take $\mL$ to be an anticode of size $A_q^*(n,d-1)$ then clearly $A_q(n,d;\mL)=1$. Using this in \eqref{eq:BE}, we get a bound
\beq  \label{eq:Del}  A_q(n,d) \leq q^n/A_q^*(n,d-1), \eeq
 known as Delsarte's code-anticode bound \cite{Delsarte_Philips}. Taking $d=xn$ where $x \in[0,1]$  we get \[ n^{-1} \log_q A_q(n,xn) \leq 1 - n^{-1} \log_q A_q^*(n,xn-1).\] Taking $\limsup_{n  \to \infty}$ we get:
 \beq \label{eq:Del1} \alpha(x) \leq 1 - \alpha^*(x), 
 \eeq
 where
 \beq \label{eq:alpha*} \alpha^*(x)= \liminf_{n \to \infty}  n^{-1} \log_q A^*_q(n, xn).\eeq
 This is the  the asymptotic form of \eqref{eq:Del}. We use the notation $B(r;n)$ and $V_q(n,r)$ to denote a Hamming ball of radius $r$ in $\mF^n$ and its volume respectively.
The ball $B(t;n)$  where   $t=\lfloor (d-1)/2 \rfloor$ in $\mF^n$  is an anticode of diameter  at most $d-1$. Let  $\mF^n =\mF^{d-1} \times  \mF^{n-d+1}$ and let $v \in \mF^{n-d+1}$ be a fixed word. Sets of the form $\mF^{d-1} \times \{v\}$ of size $q^{d-1}$ are also  anticodes of diameter  $d-1$. It follows that: 
\beq \label{eq:HS1}   \alpha^*(x) \geq \text{max}\{ H_q(x/2), x \}.\eeq
Here, we have used the well known  formula:
 \[ \label{eq:ball} \lim_{n \to \infty} n^{-1} \log_q V_q(n,t n) =  H_q(t), \;   t \in [0,\theta], \]
where,
\beq \label{eq:H}  H_q(x)  =  x \log_q(\tfrac{q-1}{x}) +(1-x) \log_q( \tfrac{1}{1-x}),  \;  x \in [0,1]. \eeq
 While the convexity of $\alpha(x)$ is an open question, it is quite easy to see that:
\begin{lemma} \label{cnvx1}  The function $\alpha^*(x)$ is $\cap$-convex. 
\end{lemma} 
 \bep
 If $S_1 \subset \mF^{n_1}$ and $S_2 \subset \mF^{n_2}$ are anticodes of diameters $d_1$ and $d_2$ respectively, then $S_1 \times S_2 \subset \mF^{n_1} \times \mF^{n_2}$ is an anticode of diameter $d_1+d_2$.
Taking $S_i$ to be $A^*_q(n_i,d_i)$  anticodes, we immediately get  
  \[A^*_q(n_1+n_2,d_1+d_2) \geq A^*_q(n_1,d_1) A^*_q(n_2,d_2).\]
 Let $n = n_1+n_2$ go to infinity with $n_1/n = t +o(1)$,  $d_1/n_1 = x+o(1)$ and $d_2/n_2 = y+o(1)$. Applying $\liminf_{n \to \infty} n^{-1} \log_q$ to this inequality we get:
\[ \alpha^*(t x + (1-t) y) \geq t \alpha^*(x) + (1-t) \alpha^*(y), \; t \in [0,1].\] 
\eep
We note that  with codes we have $d(\mC_1 \times \mC_2) = \text{min}\{d(\mC_1), d(\mC_2)\}$, which is why the above proof method does not apply to the question of convexity of $\alpha(x)$.
From \eqref{eq:HS1} and  Lemma \ref{cnvx1} we get:
\[ \alpha^*(t x + (1-t) y) \geq t H_q(x/2)+ (1-t) y, \; t \in [0,1].\] 
Let $\delta = tx+(1-t) y$. We can rewrite this as 
\[ \alpha^*(\delta) \geq f(x,y),\] 
 where $f : [0,\delta) \times (\delta,1]$ is defined by   
\beq \label{eq:f}  f(x,y)= \tfrac{y-\delta}{y-x} (H_q(x/2)-x) + \delta.\eeq
We note that 
\begin{IEEEeqnarray*}{rCl}
&\tfrac{(y-x)^2}{\delta-x} \tfrac{\partial f}{\partial y} (x,y) = H_q(x/2) -x, \\
&\tfrac{x(y-x)^2}{y(y- \delta)} \tfrac{\partial f}{\partial x} (x,y) = H_q(\tfrac{x}{2}) -x +(1-\tfrac{x}{y}) \log_q(1-\tfrac{x}{2}).
\end{IEEEeqnarray*}
There is a unique positive number  $b >0$ satisfying $H_q(b/2) = b$ (where the Hamming and Singleton bounds intersect). Therefore, 
$H_q(x/2) -x$ has the same sign as $b-x$.  Using this in \eqref{eq:f}, we see that $f(x,y) \leq \delta$ for  $x \geq b$. Therefore, in order to maximize $f(x,y)$ it suffices to consider $x <b$.
We note that    $\tfrac{\partial f}{\partial y} (x,y)$ has the same sign as $H(x/2) - x$ and hence that of $b-x$.  Since $x <b$, we see that for fixed $x<b$, the function $f(x,y)$ is maximized for $y=1$. We are now reduced to maximizing
\[ f(x,1) = 1 - (1-\delta) \tfrac{1-H_q(x/2)}{1-x},\; x \in [0,\delta].\]
\begin{lemma} \label{sp'} Let $g(x) =\tfrac{1 - H_q(x/2)}{1-x}$ for $x \in [0,1]$. 
%\begin{enumerate}
%\item  $g''(x)>0$.
 \[{\rm sign} (g'(x))  = {\rm sign}(x - \tfrac{2}{q}).\]
%\end{enumerate}
\end{lemma}
\bep We calculate:
\[   g'(x) = \tfrac{1}{2(1-x)^2} \log_q ( \tfrac{q^2x(2-x)}{4(q-1)}). \]
Therefore ${\rm sign} (g'(x)) =  {\rm sign}(\tfrac{q^2x(2-x)}{4(q-1)} - 1)$.
Next, we note that
\[ \tfrac{q^2x(2-x)}{4(q-1)} - 1 = q (x-\tfrac{2}{q}) \tfrac{(q-2) +q(1-x)}{4(q-1)} \]
has the same sign as $x-2/q$, as was to be shown. A stronger assertion is that $g(x)$ is in fact $\cup$-convex:
differentiating once more, we get:
\[ \ln(q) (1-x)^3 g''(x) = \ln(\tfrac{q^2/4}{q-1})   + (\tfrac{1}{2x-x^2} -1- \ln(\tfrac{1}{2x-x^2}))\]
We note that $q^2 \geq 4(q-1)$, and hence the first term is non-negative. The  remaining parenthetical term is non-negative using the inequality 
\beq \label{eq:log} t-1-\ln(t) \geq 0 \; \text{for }    t \geq 1, \eeq 
and the fact that $t = 1/(2x-x^2) \geq 1$ for $x \in (0,1]$. \eep
It follows from Lemma \ref{sp'} that 
 \beq  \label{eq:HSmin} \text{argmin}_{x \in [0,\delta]} \tfrac{1 - H_q(x/2)}{1-x}  = \text{min}\{\delta, 2/q\}.  \eeq
Therefore we obtain the bound:
\begin{theorem} \label{thm3} $\alpha^*(x) \geq \beta(x)$ where
\beq   \label{eq:beta}  \beta(x) = \begin{cases}
  H_q(x/2)  \!\!&\text{if }  x \in [0,2/q]\\
1 - (1-x) H_q(1)  \!\!&\text{if } x \in [2/q,1]. \end{cases}\eeq
Moreover,  $\beta(x)$ is continuously differentiable and $\cap$-convex.
\end{theorem}
We have used the relation
\beq  \label{eq:HScts}  \tfrac{1 - H_q(1/q)}{1-2/q} = H_q(1)=\log_q(q-1).  \eeq
The function $\beta(x)$ is continuously differentiable because the component for $x \geq 2/q$ is just the tangent line at $x = 2/q$ to the component for $x \leq 2/q$, i.e. to $H_q(x/2)$. We note that $\beta'(x)$ equals $H_q'(x/2)/2$ for $x \leq 2/q$ and $H_q'(1/q)/2$ for $x \geq 2/q$. Since $\beta'(x)$   is non-increasing, it follows that $\beta(x)$ is  $\cap$-convex.

In the next lemma, we show that there is a sequence of anticodes $S_n \subset \mF^n$  of diameter at most $\delta n$ such that $\lim_{n \to \infty} n^{-1} \log_q |S_n|$ equals $\beta(\delta)$, i.e. the   lower bound on $\alpha^*(\delta)$ given in theorem \ref{thm3}.
\begin{lemma} \label{aclem}
Consider the anticodes  $S(d,n)$ of diameter $d$ in $\mF^n$ (taken from \cite{AK})  given by 
\[S(d,n) = B(r_{d,n};n-d+2r_{d,n}) \times \mF^{d - 2r_{d,n}}, \, \text{where}\]
\[ r_{d,n} =  {\rm max}\{0, {\rm min}\{  \lceil \tfrac{d-1}{2} \rceil,  \lceil \tfrac{n-d-q+1}{q-2} \rceil\}\} .\] 
Then $\lim_{n \to \infty} n^{-1}\log_q |S(\delta n,n)| = \beta(\delta)$.
 \end{lemma}
\bep We note that \[ \rho=\lim_{n \to \infty} \tfrac{r_{\delta n,n}}{n} =  \begin{cases} \tfrac{\delta}{2}  \!\!&\text{if }\delta \in [0,2/q]\\
\tfrac{1- \delta}{q-2} \!\!&\text{if }\delta \in [2/q,1]. \end{cases}\]
Also  $ \lim_{n \to \infty} n^{-1} \log_q |S(\delta n, n)|$  equals
\[ (1- \delta + 2 \rho) H_q(\tfrac{\rho}{1- \delta + 2 \rho})    +(\delta  - 2\rho),\]
which simplifies to $H_q(\delta/2)$ if $\delta \leq 2/q$ and (on using \eqref{eq:HScts}) to $ 1 - (1-\delta) H_q(1)$
if $\delta \geq 2/q$. This is the same as $\beta(\delta)$.
\eep
We now have all the results we need for proving theorems \ref{HSthm} and \ref{EPthm}. However, we will state a remarkable theorem due to 
Ahlswede and Khachatrian  \cite{AK}, which we will not need.  We also obtain an asymptotic version of their result and record it as a corollary,  as it  does not seem to have appeared in literature.   In brief their theorem states that $A^*_q(n,d)$ equals $|S(d,n)|$. Moreover  any $A_q^*(n,d)$ anticode is Hamming isometric to the anticode $S(d,n)$  (with some exceptions). At the asymptotic level, the result is again remarkable: The lower bound $\beta(\delta)$ for  $\alpha^*(\delta)$ given in theorem \ref{thm3} is actually the exact value of $\alpha^*(\delta)$. Moreover  $\alpha^*(\delta)$ need not have been defined using $\liminf_{n \to \infty}$ as  $\lim_{n \to \infty} n^{-1}\log_qA_q^*(n,\delta n)$ already exists. 

\begin{theorem*} \cite{AK}  \label{AKthm}
Given   $q \geq 2$ and integers $0 \leq d \leq n$, let $r_{d,n}$ and $S(d,n)$ be as in Lemma \ref{aclem}.  Then, 
\[ A_q^*(n,d) = |S(d,n)|.\]
Moreover, up to a  Hamming isometry of $\mF^n$  an  anticode $S$ of size $A_q^*(n,d)$ must be: 
\begin{itemize}
\item  $S(d,n)$
\item  or  $S(d,n)$ with $r_{d,n}$ replaced with  $r_{d,n}-1$. This case is possible only if  $(n-d-1)/(q-2)$ is a positive  integer not exceeding $d/2$.
\end{itemize}
\end{theorem*}
\begin{corollary} \label{alpha*}
 \[ \alpha^*(x)=\begin{cases}
  H_q (x/2)  &\text{ if } 0 \leq x \leq  2/q\\
1 - (1-x) H_q(1)   &\text{ if } 2/q \leq x \leq 1 . \end{cases} \] \end{corollary}
\bep It follows from the theorem of Ahlswede and Khachatrian, together with Lemma \ref{aclem} that 
\[ \lim_{n \to \infty}  \tfrac{\log_q A_q^*(n,\delta n)}{n} = \lim_{n \to \infty} \tfrac{\log_q |S(\delta n,n)|}{ n} = \beta_q(\delta).\]
Therefore
\[ \alpha^*(\delta) = \liminf_{n \to \infty}  \tfrac{\log_q A_q^*(n,\delta n)}{n}  = \lim_{n \to \infty}  \tfrac{\log_q A_q^*(n,\delta n)}{n} = \beta(\delta).\]
\eep

\section{Proofs of theorems \ref{HSthm} and \ref{EPthm}} \label{sec3}
\subsection{Proof of Theorem \ref{HSthm}} 
If we use the bound $\alpha^*(x) \geq \beta(x)$  of  Theorem \ref{thm3} in the inequality  $\alpha(x) \leq 1 - \alpha^*(x)$ (see \eqref{eq:Del1}), we obtain the bound 
\[ \alpha(x) \leq 1 - \beta(x) =:\alpha_{HS}(x). \]
Since $\beta(x)$ is $\cap$-convex and continuously differentiable (see Theorem \ref{thm3}), it follows that  $\alpha_{HS}(x)$ is  $\cup$-convex  and continuously-differentiable. 
To show that $\alpha_{HS}(x) \leq \alpha_S(x) = 1-x$, we note that $\alpha_S(x)$ being  the secant line to $\alpha_{HS}(x)$ between $(0,\alpha_{HS}(0))$ and $(1, \alpha_{HS}(1))$,
lies above the graph of  $\alpha_{HS}(x)$ as the latter is   $\cup$-convex. To prove that $\alpha_{HS}(x)$  improves $\alpha_{H}(x)$ we note that  $\alpha_{HS}(x)$ coincides with $\alpha_H(x)$ for $x \leq 2/q$, and for $x \geq 2/q$, Lemma \ref{sp'} implies that $\alpha_H(x) \geq (1-x) H_q(1) = \alpha_{HS}(x)$.  This finishes the proof of Theorem \ref{HSthm}.\\

It is worth noting that \eqref{eq:HSmin} implies the following formula for $\alpha_{HS}(\delta)$:
\beq \label{eq:HSmin1} \alpha_{HS}(\delta) = \min\limits_{x \in [0,\delta]}   \tfrac{\alpha_{H}(x) (1 - \delta)}{1-x}.   \eeq
Since $\tfrac{\theta - \delta}{\theta-x} \leq \tfrac{1 - \delta}{1-x}$  for $x \in [0, \delta]$ and $\delta \leq \theta$,    we  get
\beq \label{eq:HSmin2} \alpha_{HS}(\delta) \geq  \alpha_{HP}(\delta): =\min\limits_{x \in [0,\delta]}   \tfrac{\alpha_{H}(x) (\theta - \delta)}{\theta-x}. \eeq
It can be shown (see subsection \ref{pf1'}) that $\alpha_{HP}(\delta)$ is an upper bound for $\alpha(\delta)$ which improves both the Hamming and Plotkin bounds.

\subsection{Proof of Theorem \ref{EPthm}} 
 It will be convenient to identify the alphabet $\mF$ with the abelian group $\bZ/q \bZ$. Given $0 \leq \delta \leq \omega$, let  $w_n = \lfloor \omega n \rfloor$ and $d_n = \lfloor \delta n \rfloor$. We take   $\mL_n \subset \mF^n$ to 
be the anticode (from  Lemma \ref{aclem}):
\beq \label{eq:1pf2} \mL_n=B(r_n;n-w_n+2r_{n}) \times \mF^{w_n - 2r_n}, \text{ where}\eeq
  \[ r_{n}={\rm max}\{0, {\rm min}\{  \lceil \tfrac{w_n-1}{2} \rceil,  \lceil \tfrac{n-w_n-q+1}{q-2} \rceil\}\}. \]
We will take the balls $B(r;m)$ to be centered at  $(0, \dots,0) \in \mF^m$.
As in Lemma \ref{aclem}, we have
\beq \label{eq:rho}   \rho :=\lim_{n \to \infty} \tfrac{r_n}{n} =   \begin{cases}
\tfrac{\omega}{2}   &\text{ if } \omega \in [0, 2/q]\\
\tfrac{1 - \omega}{q-2}   &\text{ if } \omega  \in [2/q,1] . \end{cases}\eeq
We also note that Lemma \ref{aclem} gives 
 \beq\label{eq:size_L_n} \lim_{n \to \infty} n^{-1} \log_q |\mL_n| = \beta(\omega) = 1 - \alpha_{HS}(\omega). \eeq
 Let $A_q(n,d_n;\mL_n)$ be  the maximum possible size of a code contained in $\mL_n$ and having minimum distance at least $d_n$. 
 \begin{theorem} \label{thm4} $ \lim_{n \to \infty} n^{-1}  \log_q A_q(n,d_n;\mL_n)  = 0$ if 
 \beq \label{eq:thm4}    \tfrac{\rho}{\theta(1-\omega +2 \rho)} \leq  1 -  \sqrt{\tfrac{ 1-\delta/\theta}{1- \omega + 2 \rho}}. \eeq
\end{theorem}
\bep Our proof is similar to the standard proof of the analogous result  for the Elias bound (which corresponds to taking $\rho =\omega/2$ instead of the prescription \eqref{eq:rho}).
First let $\mC \subset \mL$ be a code of size $M = A_q(n,d;\mL)$,
where $\mL \subset \mF^n$ is the anticode  $\mL=  B(r;n-w+2r) \times \mF^{w- 2r}$ 
for some $r \leq w/2$. Let 
\[ \gamma_1 (\mC)=( M  n)^{-1}  \sum_{i=1}^n \sum_{a \in \mF} m(i,a)^2, \]
where $m(i,a) = \# \{ c \in \mC : c_i = a\}$. We note that $M=\sum_{a \in \mF} m(i,a)$, and that
\[ M(M-1) d \leq \sum_{c\in \mC} \sum_{c' \in \mC} d(c,c') = n M^2(1 - \tfrac{\gamma_1}{M}).\]
We can rewrite this as:
\beq \label{eq:0thm4} M \leq \frac{d/n}{ \tfrac{\gamma_1}{M}- (1-\tfrac{d}{n})}, \; \text{provided } \tfrac{\gamma_1}{M}> 1-\tfrac{d}{n}. \eeq
For $n-w+2 r < i \leq n$   we use Cauchy-Schwarz inequality to get  $\sum_{a \in \mF} m(i,a)^2 \geq  M^2/q$. 
In particular
\beq \label{eq:1thm4}  \tfrac{1}{M^2 (w - 2r)} \sum_{i=n-w +2 r +1}^{n} \sum_{a \in \mF} m(i,a)^2 \geq \tfrac{1}{q}.\eeq
  Let $\pi_1$ be the projection of $\mF^n =   \mF^{n-w+2r} \times \mF^{w-2r}$ on to the factor  $\mF^{n-w+2r}$.
We note that for $c \in \mC$, we have wt$(\pi_1(c)) <r$ because  $\pi_1(\mC) \subset  B(r;n-w+2r)$. Here wt$(v)$ is the number of nonzero entries of $v$.
Therefore 
\[ \sum_{i=1}^{n-w+2r} \sum_{a \neq 0} m(i,a)  \leq  Mr.\]
Since $\sum_{i=1}^{n-w+2r} \sum_{a} m(i,a)  = M(n-w+2r)$ we get:
\[ S = \sum_{i=1}^{n-w+2r} m(i,0)   \geq (n-w + r )M.\]
In particular
\beq \label{eq:2thm4}  \tfrac{S}{M( n-w + 2 r)} - \tfrac{1}{q} \geq \tfrac{n-w+r}{n-w+2 r}- \tfrac{1}{q}= \theta  -\tfrac{r}{n-w+2 r}.\eeq
 We note that $\sum_{a \neq 0} m(i,a)  =  M- m(i,0)$.
By Cauchy-Schwarz inequality:
\[  \sum_{i=1}^{n-w+2r} m(i,0)^2 \geq S^2/(n-w+2 r),\; \text{ and} \]
\[  \sum_{a \neq 0} m(i,a)^2  \geq   (M - m(i,0))^2/(q-1) \]
Since  $\sum_{i=1}^{n-w+2r} \sum_{a \in \mF} m(i,a)^2$ equals 
\[  \sum_{i=1}^{n-w+2r}\left( m(i,0)^2  +  \sum_{a \neq 0}  m(i,a)^2 \right ),\]
we get:  \[\sum_{i=1}^{n-w+2r} \sum_{a \in \mF} m(i,a)^2  \geq  \!\!\! \sum_{i=1}^{n-w+2r} \!\! \left(
 \tfrac{q m(i,0)^2 + M^2  - 2 M m(i,0)}{q-1} \right)\]
 This can be rewritten as: 
% \begin{multline*}  \tfrac{1}{M^2(n-w+2r)}   \sum_{i=1}^{n-w+2r} \sum_{a \in \mF} m(i,a)^2  \geq  \\
%   \tfrac{1}{\theta}(\tfrac{S}{M(n-w+2 r)} - \tfrac{1}{q} )^2 + \tfrac{1}{q}  \end{multline*}

 \[  \tfrac{1}{M^2(n-w+2r)}   \sum_{i=1}^{n-w+2r} \sum_{a \in \mF} m(i,a)^2  \geq  
   \tfrac{1}{\theta}(\tfrac{S}{M(n-w+2 r)} - \tfrac{1}{q} )^2 + \tfrac{1}{q}  \]
 Combining this with \eqref{eq:1thm4} we get:
  \[  \tfrac{\gamma_1}{M}  \geq    \tfrac{n-w +2r}{n \theta} (\tfrac{S}{M(n-w+2r)} - q^{-1})^2 + \tfrac{1}{q}.\]
 Using \eqref{eq:2thm4} this can be written as:
 \[  \tfrac{\gamma_1}{M}  - \tfrac{1}{q}  \geq   \tfrac{n-w +2r}{n \theta} \, (\theta  -\tfrac{r}{n-w+2 r})^2.\]
Now let  $\mC_n \subset \mL_n$ be a sequence of codes of size $M_n = A_q(n,d_n;\mL_n)$.
The preceding inequality gives:
\[  \tfrac{\gamma_1(\mC_n)}{M_n}  - \tfrac{1}{q}  \geq   \tfrac{1-\omega +2 \rho}{ \theta} \, (\theta  -\tfrac{\rho}{1-\omega +2 \rho})^2 + o(1).\]
Using this in  \eqref{eq:0thm4}, we get:
\[  M_n \leq \frac{\delta+o(1)}{ \tfrac{1-\omega +2 \rho}{ \theta} \, (\theta  -\tfrac{\rho}{1-\omega +2 \rho})^2 - (\theta-\delta)+ o(1)}, \]
provided the denominator is a positive number.
Therefore,  $\lim_{n \to \infty} n^{-1}  \log_q M_n = 0$ provided
\[   \tfrac{1-\omega +2 \rho}{ \theta} \, (\theta  -\tfrac{\rho}{1-\omega +2 \rho})^2  \geq \theta  - \delta.   \]
This condition is the same as 
 \[    \tfrac{\rho}{\theta(1-\omega +2 \rho)} \leq  1 -  \sqrt{\tfrac{ 1-\delta/\theta}{1- \omega + 2 \rho}} \]
Since $M_n = A_q(n,d_n;\mL_n)$ this finishes the proof.
\eep
Using  \eqref{eq:BE} we get:
\[   \tfrac{\log_q A_q(n,d_n)}{n}  \leq 1   -    \tfrac{\log_q |\mL_n|}{n}  +  \tfrac{\log_q A_q(n,d_n;\mL_n)}{n}.\]
Taking $\limsup$ as $n \to \infty$ and using the result of Theorem \ref{thm4} and \eqref{eq:size_L_n}  we get: 
\beq \label{eq:EPmax}  \alpha(\delta) \leq \alpha_{HS}(\omega_{\text{max}}(\delta)), \eeq
where $\omega_{\text{max}}(\delta)$ is the largest value of $\omega$ for which the inequality \eqref{eq:thm4} holds. 
In order to determine $\omega_{\text{max}}(\delta)$,  we introduce functions $f_1, f_2$ on $[0,\theta]$ defined by:
\begin{IEEEeqnarray}{rcl}
 f_1(\delta) &=2 \theta(1 - \sqrt{1 - \delta/\theta})  \\
f_2(\delta)&=1 - (1 - \delta/\theta) \tfrac{(q-1)^2}{q(q-2)} 
 \end{IEEEeqnarray}

 \begin{lemma} \label{f12lem} Let $q >2$. \begin{enumerate}
\item $ f_1(\delta)  \geq f_2( \delta)$ with equality only at $\delta = \tfrac{2q-3}{q(q-1)}$.
\item  $f_2(\delta)$ is the tangent line to $f_1(\delta)$ at $\delta = \tfrac{2q-3}{q(q-1)}$.
\item $\text{sign}(f_1(\delta)- 2/q) =\text{sign}(f_2(\delta)- 2/q) = \text{sign}(\delta -\tfrac{2q-3}{q(q-1)})$.
\end{enumerate}
 \end{lemma}
 \bep  Let  $f_3(\delta):=1 -  \tfrac{q-1}{q-2} \sqrt{1 - \delta/\theta}$. We observe that
 \[ \text{sign}(f_3(\delta)) = \text{sign}(\delta -\tfrac{2q-3}{q(q-1)} ).\] 
 The  three assertions to be proved  follow respectively from the  following three relations:
 \begin{eqnarray*}
  f_1(\delta) - f_2(\delta)  &=&   (1 - 2/q) \,f_3(\delta)^2, \\
  f_1'(\delta) - f_2'(\delta)&=&f_3(\delta)/\sqrt{1 - \delta/\theta},\\
 \tfrac{ f_1(\delta) - 2/q}{2(1 - 2/q)} =  \tfrac{ f_2(\delta) - 2/q}{(1-2/q) +   \theta  \sqrt{1 - \delta/\theta}}&=& f_3(\delta) .
 \end{eqnarray*}
  \eep 

 \begin{proposition} \label{EPprop}  \[ \omega_{\text{max}}(\delta) = \begin{cases}  2 \theta(1 - \sqrt{1 - \delta/\theta})    &\text{ if }  \delta \in [0,\tfrac{2q-3}{q(q-1)}] \\ 1 - (1 - \delta/\theta) \tfrac{(q-1)^2}{q(q-2)}  &\text{ if } \delta \in [\tfrac{2q-3}{q(q-1)},1].   \end{cases}\]
The function  $\omega_{\text{max}}(\delta)$  is  increasing, continuously differentiable, and $\cup$-convex  on $[0, \theta]$.
  \end{proposition}
\bep The inequality \eqref{eq:thm4}   reduces to 
\[ \omega \leq \begin{cases} f_1(\delta) &\text{ if  $\rho=\omega/2$}\\
f_2(\delta) &\text{ if  $\rho=(1-\omega)/(2-q)$}, \end{cases} \]
where $\rho$ is as given in  \eqref{eq:rho}. Therefore, for a given $\delta \in [0,\theta]$, the quantity $\omega_{\text{max}}(\delta)$  is the maximum element of the set
\[ \{ \omega: \delta \leq \omega \leq \text{min}\{f_1(\delta), 2/q\}   \}      \cup \{ \omega: \text{max}\{\delta, 2/q \} \leq \omega \leq f_2(\delta) \}. \]
If $\delta  \geq \tfrac{2q-3}{q(q-1)}$, then $f_2(\delta) \geq 2/q$ (by Lemma \ref{f12lem}) and hence, the maximum of this set is  $f_2(\delta)$.
If $\delta  \leq \tfrac{2q-3}{q(q-1)}$, then $f_2(\delta) \leq f_1(\delta) \leq 2/q$ (by Lemma \ref{f12lem}) and hence, the maximum of this set is  $f_1(\delta)$. This proves the asserted formula for $\omega_{\text{max}}(\delta)$.\\

We note that the second component of $\omega_{\text{max}}(\delta)$ is the tangent line to the first component at $x = \tfrac{2q-3}{q(q-1)}$. Therefore  $\omega_{\text{max}}(\delta)$ is continuously differentiable. The derivative of  $\omega_{\text{max}}(x)$ is $1/ \sqrt{1 - x/\theta}$ for $x \leq \tfrac{2q-3}{q(q-1)}$, and constant at $\tfrac{q-1}{q-2}$ for $x \geq 
\tfrac{2q-3}{q(q-1)}$.  Since the derivative is positive, the function is increasing. Since the derivative is non-decreasing, we see that the function is $\cup$-convex.
\eep

\emph{Proof of $\alpha_{EP}$ being an upper bound}:  We note from lemma \ref{f12lem} that \[ \text{sign}(\omega_{\text{max}}(\delta) - 2/q) = \text{sign}(\delta - \tfrac{2q-3}{q(q-1)}).\]
Therefore $\alpha_{HS}(\omega_{\text{max}}(\delta))$ is just the function $\alpha_{EP}(\delta)$ defined in of theorem \ref{EPthm}.  The bound $\alpha(x) \leq \alpha_{EP}(x)$  now follows from  \eqref{eq:EPmax}. \\  

\emph{Proof of $\alpha_{EP}$ being continuously differentiable}: The function $\alpha_{EP}(x) = \alpha_{HS}(\omega_{\text{max}}(x))$ being a composition of continuously differentiable functions, is itself continuously differentiable. \\

\emph{Proof of $\alpha_{EP}$ being $\cup$-convex}:
Both  the functions $\alpha_{HS}$ and $\omega_{\text{max}}$ are $\cup$-convex, but $\alpha_{HS}$ is decreasing and hence it is not obvious that $\alpha_{EP}(x) = \alpha_{HS}(\omega_{\text{max}}(x))$ is $\cup$-convex. We will show instead that the derivative $\alpha_{EP}'$ is non-decreasing. Since $\alpha_{EP}'$ is constant for $x \geq \tfrac{2q-3}{q(q-1)}$, it suffices to show that $\alpha_E''(x) >0$ for $x \in (0, \tfrac{2q-3}{q(q-1)}]$.  This follows from the next lemma.
\begin{lemma} \label{Econv} The Elias bound $\alpha_E(x)$ is $\cup$-convex on $[0,\delta_E]$ and $\cap$-convex on $[\delta_E,\theta]$ where $\delta_E$ satisfies:
\[ \tfrac{2q-3}{q(q-1)} < \delta_E < \tfrac{3}{4} (\tfrac{q - 4/3}{q-1}).\]
\end{lemma}
\bep  Let $Z(x) =  \theta(1 - \sqrt{1 - x/\theta})$. A calculation shows that
\[ 4 \theta \ln(q) (1 - \tfrac{Z(x)}{\theta})^3 \alpha_E''(x) = \varphi(Z(x)), \text{ where}\]
\[ \varphi(z) =  \int_{\tfrac{1-\theta}{1-z}}^{\tfrac{\theta}{z}} (1 - 1/t) dt.\]
To see this, we note: $\alpha_E(x) = H_q(Z(x))$ and hence 
\[  \ln(q)  \alpha_E''(x)  = \tfrac{Z'}{Z(1-Z)} +Z''  \ln(\tfrac{(q-1)(1-Z)}{Z}).\]
Since $Z' = 1/(2 (1  - Z/\theta))$ and $Z'' = Z'/(2 \theta (1 - Z/\theta)^2)$, we get 
\[ 4 \theta \ln(q) (1 - \tfrac{Z(x)}{\theta})^3 \alpha_E''(x) =\int_{\tfrac{1-\theta}{1-Z(x)}}^{\tfrac{\theta}{Z(x)}} (1 - 1/t) dt,\]
as desired. It follows that sign$(\alpha_E''(x)) = \text{sign}(\varphi(Z(x))$. Next we note that $Z(x)$ is   increasing on $[0,\theta]$ and  
\[ Z(\tfrac{2q-3}{q(q-1)}) = 1/q, \quad Z(\tfrac{3}{4} (\tfrac{q - 4/3}{q-1})) = 1/2.\]
It now suffices to show that 
\[ \text{sign}(\varphi(z)) = \text{sign}(z_E - z), \text{  for some } z_E \in (\tfrac{1}{q},\tfrac{1}{2}).\]
We note that
 \[ \varphi'(z) = (z - \tfrac{1}{2})  \, \tfrac{2(\theta-z)}{z^2(1-z)^2}.\]
Thus $\varphi(z)$ is decreasing on $[0,1/2]$ and increasing on $[1/2,\theta]$. 
In order to show sign$(\varphi(z)) = \text{sign}(z_E - z)$ for some $z_E \in (1/q,1/2)$, it suffices to show that $\varphi(1/q) >0$ and $\varphi(1/2) <0$. We calculate
 \[ \tfrac{1}{2} \varphi(1/q) = (\tfrac{q-1}{2} -1 -\ln(\tfrac{q-1}{2})  ) +  (\tfrac{2q-3}{2q-2} - \ln(2) ).\]
Since $q \geq 3$, we have  $\tfrac{q-1}{2} \geq 1$. The  inequality \eqref{eq:log}  implies that the first parenthetical term above is non-negative.
Again $q \geq 3$ implies \[\tfrac{2q-3}{2q-2} - \ln(2) \geq \tfrac{3}{4}-\ln(2) >0,\]  and hence the second parenthetical term is positive. Thus $\varphi(1/q) >0$.

Next, we note that $\varphi(1/2) = 2 - 4/q - \ln(q-1)$. The function $a(t) = 2 - 4/t-\ln(t-1)$ satisfies
\[ a'(t) = - \tfrac{(t-2)^2}{t^2(t-1)},\] 
and $a(3) = 2/3 - \ln(2) <0$. Therefore $a(t) <0$ for $t \geq 3$, and hence $\varphi(1/2) <0$ for all $q \geq 3$. \\
\eep

\emph{Proof that  $\alpha_{EP}$ improves the Plotkin bound}:
We have already shown that $\alpha_{EP}(x)$ is $\cup$-convex, and hence $\alpha_{EP}(x)$ lies below the secant line between $x=0$ and $x = \theta$, which  is the Plotkin bound.\\

\emph{Proof that  $\alpha_{EP}$ improves the Elias bound}:  This does not readily follow from our results thus far, and requires more work.  The 
 characterization of $\alpha_{EP}(x)$ given in the next theorem clearly implies $\alpha_{EP}(x) \leq \alpha_E(x)$.
\begin{theorem} \label{EPchar}
$\alpha_{EP}(\delta) = \min\limits_{x \in [0,\delta]}   \tfrac{\alpha_{E}(x) (\theta - \delta)}{\theta-x}.$ 
\end{theorem}
\bep  The theorem immediately follows if we show that $\tfrac{\alpha_E(x)}{\theta-x}$ is decreasing on $[0, \tfrac{2q-3}{q(q-1)}]$ and increasing on $[\tfrac{2q-3}{q(q-1)}, \theta]$.
We will use the notation from the proof of Lemma \ref{Econv}.
Since $\alpha_E(x)$ is $\cap$-convex for $x \geq \delta_E$, it follows that the slope  $\alpha_E(x)/(\theta-x)$ of the secant between $x$ and $\theta$ is increasing.
It remains to show that $\tfrac{\alpha_E(x)}{\theta-x}$ is decreasing on $[0, \tfrac{2q-3}{q(q-1)}]$ and increasing on $[\tfrac{2q-3}{q(q-1)}, \delta_E]$.
Since $Z(x)$ is an increasing function, with $Z(0)=0,Z(\tfrac{2q-3}{q(q-1)})=1/q$, and $(1-x/\theta) = (1 - z/\theta)^2$,  it suffices to show that
\[ h(z) =\tfrac{1 - H_q(z)}{(\theta-z)^2}, \]
 is decreasing on $[0, 1/q]$ and increasing on $[1/q, z_E]$ where $z_E = Z(\delta_E)$. A calculation shows that
 \[ \ln(q) (\theta-z)^3  h'(z) = \int_{1/q}^{z} \varphi(t) dt.\]
 To see this we note that either side of this equation evaluates to $(\theta+z) \ln( \frac{z}{(1-z)(q-1)}) + 2 \ln(q(1-z))$.
 Since $\varphi(t) >0$ for $t\in (0,z_E)$, we see
 \[ \text{sign}(h'(z)) = \text{sign}(z - 1/q),\; z \in [0,z_E].\]
 Thus we have also shown that $h(z)$ is decreasing for $z \in [0,1/q]$ and increasing on $[1/q, z_E]$ as required.
 \eep
The bounds $\alpha_{HS}$, $\alpha_{HP}$ and $\alpha_{EP}$ are related as 
\[ \alpha_{EP}(\delta) \leq \alpha_{HP}(\delta) \leq \alpha_{HS}(\delta).\]
We have already shown  $\alpha_{HP}(\delta) \leq \alpha_{HS}(\delta)$ in \eqref{eq:HSmin2}.  Since $\alpha_E(x) \leq \alpha_H(x)$ for all $x$, we note that
\[ \min\limits_{x \in [0,\delta]}   \tfrac{\alpha_{E}(x) (\theta - \delta)}{\theta-x} \leq \min\limits_{x \in [0,\delta]}   \tfrac{\alpha_{H}(x) (\theta - \delta)}{\theta-x}.\]
Thus $\alpha_{EP}(\delta)  \leq \alpha_{HP}(\delta)$. We end this section with a plot comparing $\alpha_E(x)$ and $\alpha_{EP}(x)$ for $q=16$.
\begin{figure}[!t]
\centering
\includegraphics[width=3.5in]{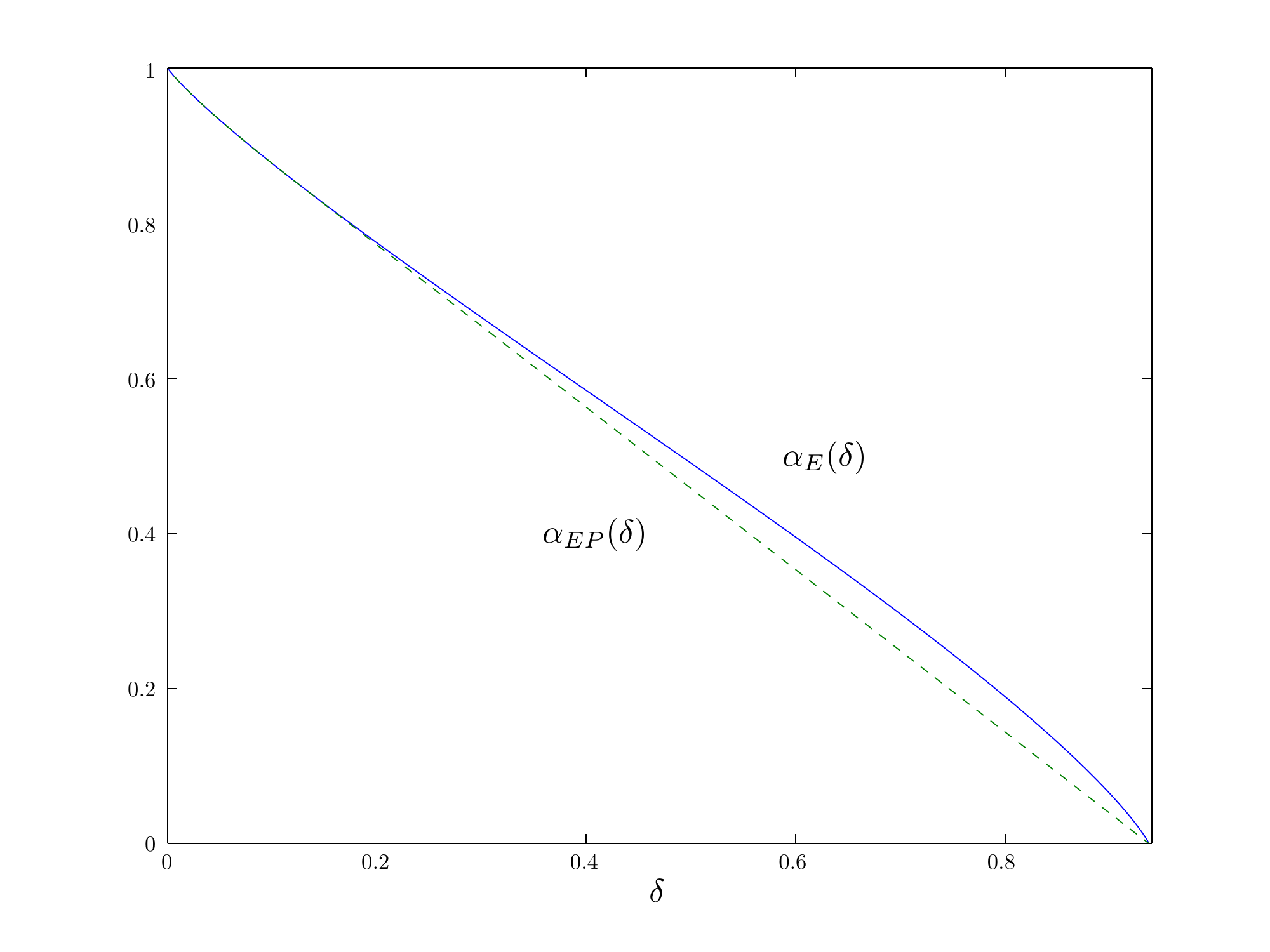}
\caption{$\alpha_E(\delta)$ and $\alpha_{EP}(\delta)$ for $q = 16$.}
\label{fig_sim}
\end{figure}

\subsection{Another proof of Theorem \ref{HSthm}} \label{pf1'}
Another proof of  $\alpha_{HS}(x)$ being an upper bound for $\alpha(x)$  can be given using the following theorem of Laihonen and Litsyn:
\begin{theorem*} \cite{LL} \label{ac_thm} 
Let $\delta_1, \delta_2, \mu \in [0,1]$.
\beq \label{eq:ac_thm} \alpha((1-\mu) \delta_1 + \mu \delta_2) \leq  (1-\mu) \alpha_{H}( \delta_1) + \mu \alpha(\delta_2).  \eeq
\end{theorem*}
\bep We give a quick proof.
The result follows  from the inequality:
\[ A_q(n_1+n_2,d_1+d_2) \leq \frac{q^{n_1} A_q(n_2,d_2)}{V_q(n_1,d_1/2)},\]
by taking $n_1+n_2 = n  \to \infty$, and $n_1/n, n_2/n, d_1/n_1$  and $d_2/n_2$ going to  $1- \mu, \mu, \delta_1$ and $\delta_2$ respectively.  
 The above inequality in turn comes from the Bassalygo-Elias lemma \eqref{eq:BE} 
\[ A_q(n_1+n_2,d_1+d_2) \leq \frac{q^ {n_1+n_2} A_q(n_1+n_2,d_1+d_2;\mL)}{|\mL|},\] by taking $\mL = B(d_1/2;n_1) \times \mF^{n_2}$, and observing that  $A_q(n_1+n_2,d_1+d_2;\mL) \leq  A_q(n_2,d_2)$.  (If $\mC$ is a $A_q(n_1+n_2,d_1+d_2;\mL)$ code, and $\pi_2: \mF^{n_1} \times \mF^{n_2} \to \mF^{n_2}$ is the projection on the second factor, then 
the restriction of $\pi_2$ to $\mC$ is injective, and $\pi_2(\mC)$ has minimum distance at least $d_2$.)

\eep
If we set $\delta_2=1$,   $\delta_1=x$ and $\mu =(\delta - \delta_1)/(1 - \delta_1)$ in \eqref{eq:ac_thm}, we get:
  \[    \frac{\alpha(y)}{1- y}  \leq  \frac{\alpha_{H}(x)}{1-x} \; \text{ for }\,  x \leq y .\]
 Thus, 
 \[ \alpha(\delta) \leq \min\limits_{x \in [0,\delta]}   \tfrac{\alpha_{H}(x) (1 - \delta)}{1-x} = \alpha_{HS}(\delta),\]
where we have used  \eqref{eq:HSmin1}. \\

We now prove that the function $\alpha_{HP}(\delta)$ defined in  \eqref{eq:HSmin2} is an upper bound for $\alpha(\delta)$.
Taking $\delta_2=\theta$, $\delta_1=x$,  and $\mu = (y - \delta_1)/(\theta - \delta_1)$ in \eqref{eq:ac_thm}, we get:
  \[  \frac{\alpha(y)}{\theta- y}  \leq  \frac{\alpha_{H}(x)}{\theta-x} \; \text{ for }\,  x \leq y \]
 Thus 
 \beq \label{eq:HP}  \alpha(\delta) \leq \min\limits_{x \in [0,\delta]}   \tfrac{\alpha_{H}(x) (\theta - \delta)}{\theta-x} = \alpha_{HP}(\delta).\eeq

It is not known if the inequality \eqref{eq:ac_thm} (the theorem of Laihonen-Litsyn) holds if we replace $\alpha_H$ by $\alpha_E$. If such a result were true, then 
the derivation of the bound $\alpha_{HP}(x)$ above with $\alpha_H$ replaced with $\alpha_E$ would immediately yield Theorem \ref{EPchar}. 
%We elaborate on  this in the next section (see \eqref{eq:imp1}).   
We believe that such an  inequality
\beq \label{eq:E_ineq} \alpha((1-\mu) \delta_1 + \mu \delta_2) \leq  (1-\mu) \alpha_{E}( \delta_1) + \mu \alpha(\delta_2), \eeq
  must be true (it would surely be true if $\alpha(x)$ is $\cup$-convex), but we believe it cannot be obtained just by a simple application of the  Bassalygo-Elias lemma  \eqref{eq:BE}. 
 If \eqref{eq:E_ineq}  holds, we can obtain an upper bound which improves the Laihonen-Litsyn bound \cite{LL}. We recall  that the Laihonen-Litsyn bound, which we denote $\alpha_{HMRRW}$ is a hybrid of the Hamming and MRRW bounds. It coincides with the Hamming bound for $\delta \in [0,a]$ and with the MRRW bound for $[b,\theta]$  where $a<b$ are points such that the straight line joining $(a,\alpha_H(a))$ and $(b,\alpha_{MRRW}(b))$ is a common tangent to both $\alpha_H$ at $a$  and $\alpha_{MRRW}$ at $b$. Since the Hamming bound is good for small $\delta$ and the MRRW bound good for large $\delta$, the Laihonen-Litsyn bound combines the best features of both bounds in to a single bound. To obtain this bound, we note that \eqref{eq:ac_thm} implies the inequality 
 \[ \alpha((1-\mu) \delta_1 + \mu \delta_2) \leq  (1-\mu) \alpha_{H}( \delta_1) + \mu \alpha_{MRRW}(\delta_2).\]
 We  fix $\delta = (1-\mu) \delta_1+ \mu \delta_2$ and choose $\delta_1$ and $\delta_2$ optimally in order to minimize 
the right hand side.  This yields the $\alpha_{HMRRW}$ bound.  Since the second MRRW bound $\alpha_{MRRW2}$ improves the first MRRW bound $\alpha_{MRRW}$,  a better version $\alpha_{HMRRW2}$ of the Laihonen-Litsyn bound (see \cite[Theorem 2]{BL}) can be obtained by using $\alpha_{MRRW2}$ in place of $\alpha_{MRRW2}$. Since the Elias bound $\alpha_E(\delta)$ is better than the Hamming bound $\alpha_H(\delta)$ for all $\delta$, in case \eqref{eq:E_ineq} is true, repeating this procedure with $\alpha_E$ replacing $\alpha_H$, would yield the hybrid Elias-MRRW bounds $\alpha_{EMRRW}(\delta), \alpha_{EMRRW2}(\delta)$ which would improve the respective Laihonen-Litsyn bounds $\alpha_{HMRRW}, \alpha_{HMRRW2}(\delta)$. We leave the  question of the truth of \eqref{eq:E_ineq} open.  
\section{On the  convexity of $\alpha(x)$} \label{convx}
A fundamental open question  about the function $\alpha(x)$ is whether it is $\cup$-convex. In other words is it true that
\beq \label{eq:conv} \alpha((1-t) x + t y ) \leq (1-t) \alpha(x) + t \alpha(y), \; t \in [0,1].\eeq
It is worth noting that non-convex upper bounds like the Elias bound and the MRRW bound admit corrections to the non-convex part:  the bound $\alpha_{EP}(x)$ for the Elias bound and the Aaltonen straight-line bound (see the theorem below and the Appendix) for the MRRW bound.  This may be viewed as some kind of evidence supporting the truth of \eqref{eq:conv}.
It is known that \eqref{eq:conv} holds for $x=0$ (for example by taking $\delta_1=0$ in \eqref{eq:ac_thm}). Another way to state this is that 
\[ (1-\alpha(x))/x\, \text{ is decreasing on } [0, 1]. \]
As a consequence,  if $\alpha_u(x)$ is any upper bound  for $\alpha(x)$ we obtain a better upper bound
\beq \label{eq:Aalt1}  \alpha(\delta) \leq  \tilde\alpha_u(\delta) =  1 - \max_{x \in [\delta,\theta]} \frac{(1 -  \alpha_u(x)) \delta}{x} \eeq
To see this we use:
\[ \tfrac{1-\alpha(\delta)}{\delta}  \geq \tfrac{1-\alpha(x)}{x} \geq \tfrac{1-\alpha_u(x)}{x},  \text{for }  x \in [\delta,\theta].\]
Thus $\tfrac{1-\alpha(\delta)}{\delta}  \geq \max_{x \in [\delta,\theta]} \frac{(1 -  \alpha_u(x))}{x}$ as desired.
If $(1-\alpha_u(x))/x$ is a decreasing function then the improved bound $\tilde \alpha_u(x)$ coincides with $\alpha_u(x)$, but otherwise $\tilde \alpha_u(x)$ improves $\alpha_u(x)$.
For example let $\alpha_u(x)$  be the first MRRW bound $\alpha_{MRRW}(x)=$
\[  H_q( (   \sqrt{\theta (1-x)}  - \sqrt{ x (1-\theta) })^2), \,  x \in [0,\theta]. \]
It can be shown that  that $(1-\alpha_{MRRW}(x))/x$ fails to be decreasing near $x=0$, and similarly $\alpha_{MRRW}(x)$ fails to be $\cup$-convex near $x=0$.
This is immediately rectified by passing to the improved bound $\tilde \alpha_{MRRW}(x)$, resulting in the following theorem of Aaltonen.
 \begin{theorem*}  (Aaltonen bound) \cite{Aalt} \cite[p.53]{Tsfasman} Let $q>2$.
$\alpha(x) \leq \tilde\alpha_{MRRW}(x)$ where 
\beq \label{eq:MRRW1} \tilde\alpha_{MRRW}(x) =  \begin{cases} 
1 - \tfrac{x H_q(1)}{1 - 2/q} &\text{ if } x \in [0, (1 - \tfrac{2}{q})^2] \\
\alpha_{MRRW}(x)  &\text{ if }  x \in [ (1 - \tfrac{2}{q})^2, \theta]
 \end{cases} \eeq
 This bound is $\cup$-convex, continuously differentiable,  and improves the MRRW bound.
\end{theorem*}
We note that for $x \leq  (1 - 2/q)^2$  the bound $\tilde \alpha_{MRRW}(x)$ coincides with the tangent line to $\alpha_{MRRW}(x)$ at $(1 - 2/q)^2$. In particular 
$\tilde \alpha_{MRRW}(x)$ is continuously differentiable. The assertion that $\tilde \alpha_{MRRW}(x)$ improves $\alpha_{MRRW}(x)$ follows from the fact that $\tilde \alpha_u(x) \leq \alpha_u(x)$ for any upper bound $\alpha_u(x)$ for $\alpha(x)$. The other assertions are proved in the appendix.\\

On the other hand, it is not known if the convexity condition \eqref{eq:conv} holds for $y= \theta$, in other words if $\alpha(x)/(\theta-x)$ is a decreasing function of $x$. We conjecture that this is true (see Conjecture \ref{conj1}). As evidence for this conjecture, we now show that the bounds $\alpha_{EP}$, $\alpha_{HP}$ and $\alpha_{HS}$ can be obtained without doing any work, if we assume the truth of Conjecture \ref{conj1}:  if $\alpha_u(x)$ is any upper bound  for $\alpha(x)$ we obtain a better upper bound
\beq \label{eq:imp1}  \alpha(\delta) \leq \alpha_u^ \dagger(\delta):=   \min_{x \in [0,\delta]}  \frac{\alpha_u(x)( \theta - \delta)}{\theta - x} \eeq
To see this we use:
\[ \tfrac{\alpha(\delta)}{\theta - \delta}  \leq \tfrac{\alpha(x)}{\theta-x} \leq \tfrac{\alpha_u(x)}{\theta-x}\;  \text{  for }  x \in [0,\delta].\]
Thus $\alpha(\delta)  \leq      \min_{x \in [0,\delta]}  \frac{\alpha_u(x)( \theta - \delta)}{\theta - x}  $ as desired.
In case  $\tfrac{\alpha_u(x)}{\theta-x} $ is a decreasing function then the improved bound $\alpha_u^ \dagger(x)$ coincides with $\alpha_u(x)$, but otherwise $\alpha_u^ \dagger(x)$ improves $\alpha_u(x)$.  Taking $\alpha_u(x)$ to be the Elias bound, we get $\alpha_u^\dagger(x)$ to be the bound $\alpha_{EP}$. This is the content of Theorem \ref{EPchar}.
Taking $\alpha_u(x)$ to be the Hamming bound, we get $\alpha_u^\dagger(x)$ to be the bound $\alpha_{HP}$. This is the content of \eqref{eq:HP}.
Moreover, if $\alpha(x)/(\theta-x)$ is decreasing then  $\alpha(x)/(1-x)$  being the product of the non-negative decreasing functions  $\alpha(x)/(\theta-x)$ and $(\theta-x)/(1-x)$
is itself decreasing. Thus we obtain $\alpha(\delta)  \leq      \min_{x \in [0,\delta]}  \frac{\alpha_u(x)( 1 - \delta)}{1 - x}$. Taking $\alpha_u(x)$ to be the Hamming bound,  the bound $\min_{x \in [0,\delta]}  \frac{\alpha_u(x)( 1 - \delta)}{1 - x}$ is $\alpha_{HS}(\delta)$. This is the content of \eqref{eq:HSmin1}.
\appendices \label{App1}
\section{Aaltonen's straight-line bound}
The bound $\tilde \alpha_{MRRW}$  presented above was obtained by Aaltonen in \cite[p.156]{Aalt}.  The bound follows from \eqref{eq:Aalt1} and the following result
 \beq  \label{eq:MRRWmin} \text{argmax}_{x \in [\delta,\theta]} \tfrac{1 - \alpha_{MRRW}(x)}{x}  = \text{max}\{\delta, (1 - 2/q)^2\}.  \eeq
The argmax above is not straightforward to obtain, and to quote from \cite{Aalt}, was found by a mere chance. The derivation is not presented in \cite{Aalt}. The purpose of this appendix is to i) record a proof of \eqref{eq:MRRWmin}, and ii) to  prove that $\tilde \alpha_{MRRW}(x)$ is $\cup$-convex. The author thanks 
Tero Laihonen for providing a copy of Aaltonen's work \cite{Aalt}, which is not easily available.\\

Let $\xi:[0, \theta] \to [0,\theta]$ be the function defined by $\xi(x) = (   \sqrt{\theta (1-x)}  - \sqrt{ x (1-\theta) })^2$. We note that  $\alpha_{MRRW}(x) = H_q(\xi(x))$,  and that  $\xi(x)$ decreases from $\theta$ to $0$  as $x$ runs from $0$ to $\theta$.   It is easy to check that $\xi( \xi (x)) = x$  for $x \in [0,\theta]$.
Therefore we can invert the relation $y = \xi(x)$ as $x = \xi(y)$. We also note that $\xi((1 - 2/q)^2) = 1/q$. Therefore \eqref{eq:MRRWmin} is equivalent to the assertion:
 \beq  \label{eq:MRRWmin1} \text{argmax}_{y \in [0,t]} \tfrac{1 - H_q(y)}{\xi(y)}  = \text{min}\{t, 1/q\}.  \eeq
In terms of  $h_A(y) := \tfrac{1 - H_q(y)}{\xi(y)}$ we must show 
\[ \text{sign}(h'_A(y)) = \text{sign}(1/q-y), \quad y \in (0,\theta).\]
A calculation shows that:
\[ h_A'(y) \xi(y)^{3/2} \sqrt{\tfrac{y(1-y)}{\theta(1-\theta)}} \ln(q) =  \sqrt{\tfrac{y}{1- \theta}} \,  \ln(\tfrac{y}{\theta}) + \sqrt{\tfrac{1- y}{\theta}} \,  \ln(\tfrac{1-y}{1-\theta}):= G(y) \]
Clearly $\text{sign}(h'_A(y)) = \text{sign}(G(y))$. Therefore,  we must show that $\text{sign}(G(y))= \text{sign}(1/q-y)$ for $y \in (0,\theta)$.  Clearly $G(1/q)=G(1-\theta)=0$. First we will prove that  $G(y) >0$ on $[0,1/q)$. We calculate:
\[ -\sqrt{\theta(1-\theta)}\,  G'(y)  = \int^{ \sqrt{\tfrac{\theta}{y}}}_{ \sqrt{\tfrac{1-\theta}{1-y}}} \ln(t) dt   \\=  \int^{1}_{ \sqrt{\tfrac{1-\theta}{1-y}}} \ln(t) dt  + \int^{ \sqrt{\tfrac{\theta}{y}}}_{1} \ln(t) dt.\]
We make the substitution $t= 1/\tau$ in  the first integral to obtain: 
\[ -\sqrt{\theta(1-\theta)}\, G'(y)  = \int^{\sqrt{\tfrac{1-y}{1-\theta}}}_{1} \ln(t) (1 - \tfrac{1}{t^2})  dt  +\int^{ \sqrt{\tfrac{\theta}{y}}}_{ \sqrt{\tfrac{1-y}{1-\theta}}} \ln(t) dt. \]
  We note that $t \geq 1$ in both the integrals,  and hence both the integrands are non-negative. Consequently, the first integral is positive, and the second integral is also positive
 when $\sqrt{\theta/y} > \sqrt{(1-y)/(1-\theta)}$. For $y \in [0,\theta]$,
 this inequality is equivalent  to $(\theta-y)(1-\theta-y) > 0$ which in turn is equivalent to $y < 1-\theta$ i.e. $y \in [0,1/q)$.  Thus, for $y \in (0,1/q)$, we have shown that $G'(y)< 0$.
 Since $G(0) =  \ln(q)  \sqrt{q/(q-1)} >0$ and $G(1/q)=0$, the fact that  $G(y)$ is strictly decreasing on $[0,1/q]$ implies   $G(y) >0$ on $[0,1/q)$. \\

Next we prove $G(y) < 0$ on $(1/q,\theta)$. Differentiating the expression for $G'(y)$ we  get:
\[ 4 \sqrt{\theta(1- \theta)} \,G''(y)   = 
\ln (\tfrac{\theta}{y})  \tfrac{\sqrt{\theta}}{y^{3/2}} +  \ln (\tfrac{1-\theta}{1-y})  \tfrac{\sqrt{1-\theta}}{(1-y)^{3/2}}.
\]
Differentiating once more, we get:
\[ 4 \sqrt{\theta(1- \theta)} \, G'''(y)   =  
 \tfrac{\sqrt{1-\theta}}{(1-y)^{5/2}}    (1+ \tfrac{3}{2}  \ln(\tfrac{1-\theta}{1-y}))
- \tfrac{\sqrt{\theta}}{y^{5/2}}(1+ \tfrac{3}{2} \ln (\tfrac{\theta}{y}))  
 \]
 The second term  $- \tfrac{\sqrt{\theta}}{y^{5/2}}(1+ \tfrac{3}{2} \ln (\tfrac{\theta}{y}))$ 
is negative on $[1/q, \theta)$ because   $\theta/y >1$ on this interval. The first term 
$ \tfrac{\sqrt{1-\theta}}{(1-y)^{5/2}}  (1+ \tfrac{3}{2}  \ln(\tfrac{1-\theta}{1-y}))$ has the same sign as $y -(1 - q^{-1}e^{2/3})$ for $y \in [1/q, \theta)$. It follows that $G'''(y)< 0$ for $y \in [1/q, 
1 - q^{-1}e^{2/3}]$. (We note that the condition for $1/q <   1 - q^{-1}e^{2/3}$, is $q \geq 3$, which is the case here).

\begin{figure}[!t]
\centering
\includegraphics[width=3.5in]{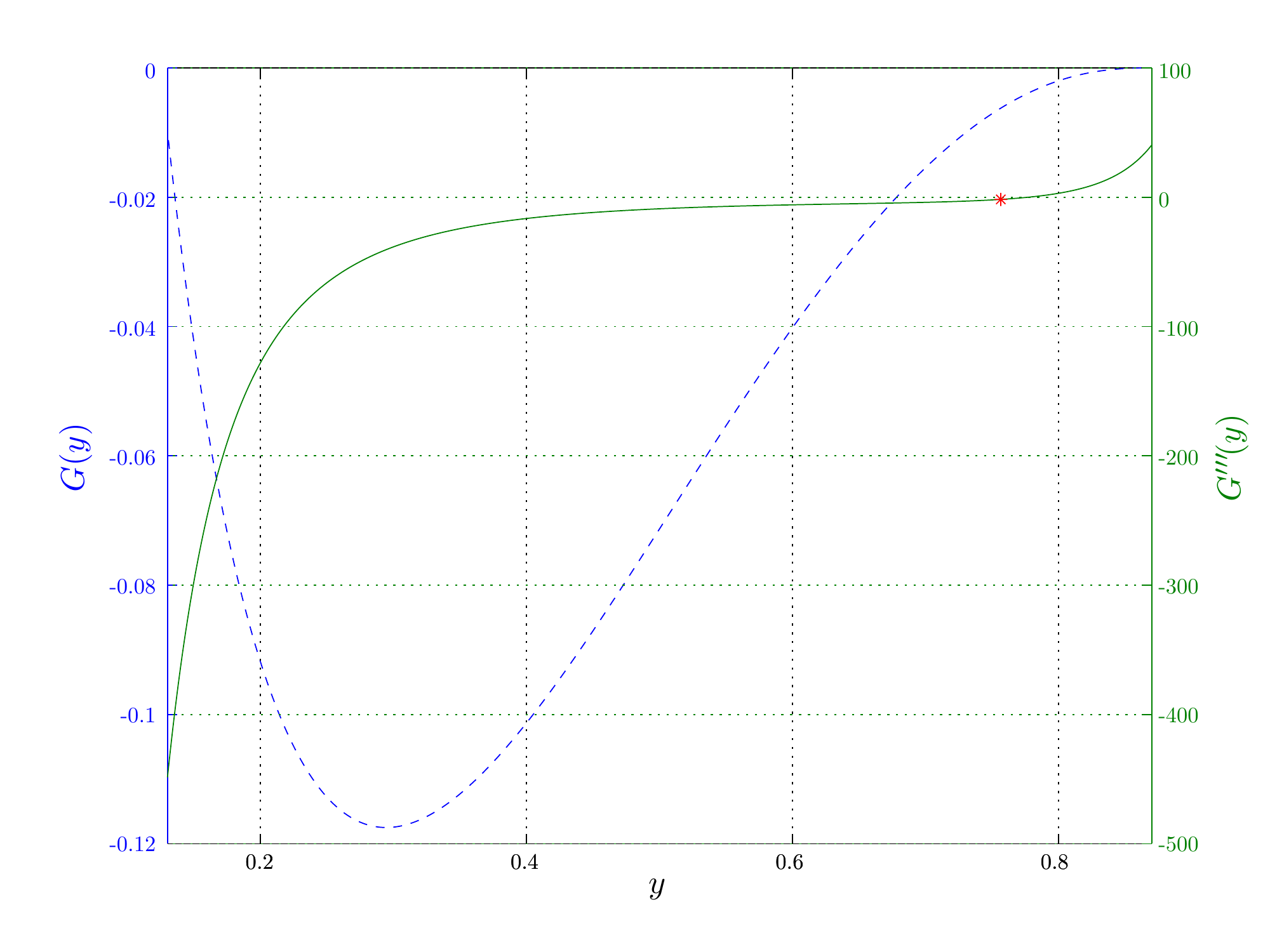}
\caption{Graphs of $G(y)$ and $G'''(y)$ on $[1/q,\theta]$  for $q = 8$.}
\label{fig:Gfunc}
\end{figure}

 For  $y \in (1 - q^{-1}e^{2/3}, \theta)$,  as above $\tfrac{\sqrt{1-\theta}}{(1-y)^{5/2}} (1+ \tfrac{3}{2}  \ln(\tfrac{1-\theta}{1-y}))$  is positive.  It is also an  increasing function of $y$, because $(1-\theta)/(1-y)$  increases with $y$.  For $y \in [1 - q^{-1}e^{2/3}, \theta]$, we note that  $\theta/y$ decreases with $y$ and $\theta/y \geq1$.  Therefore  the term $- \tfrac{\sqrt{\theta}}{y^{5/2}}(1+ \tfrac{3}{2} \ln (\tfrac{\theta}{y}))$   increases with $y$. 
Thus $G'''(y)$ is an increasing function of $y$ for $y \in [1 - q^{-1}e^{2/3}, \theta]$.
We note the boundary conditions on $G'''(y)$: we have $G'''(1/q) <0 < G'''(\theta)$. To see this we note that   
\[ -4 (\theta(1-\theta))^3 G'''(1/q) =   \tfrac{3}{2}  \ln(q-1) ( \theta^3+(1-\theta)^3) + ( \theta^3- (1-\theta)^3) >0 \]
because $q > 2$  is equivalent to   $\theta > 1- \theta$  as well as $\ln(q-1) >0$. Also 
\[ G'''(\theta)= \tfrac{2 \theta-1}{4 (\theta(1- \theta))^{2.5}} > 0.\]
Since $G'''(1 - q^{-1}e^{2/3})<0 < G'''(\theta)$ and $G'''(y)$ is increasing on 
$[1 - q^{-1}e^{2/3}, \theta]$, we conclude that 
there is a unique $y_0$ in the interior of this interval such that $G'''(y)$ has the same sign as $y-y_0$ on this interval. Together with the fact that $G'''(y) <0$ on $[1/q, 1 - q^{-1}e^{2/3}]$, we obtain:
\[ \text{sign}(G'''(y)) =  \text{sign}(y-y_0) \quad \text{ on } \; [1/q,\theta].\]
This is illustrated in Figure \ref{fig:Gfunc},  which shows the graphs of $G(y)$ (dashed plot) and $G'''(y)$  on $[1/q, \theta]$ for $q=8$. The point $(y,G'''(y))$ for $y=1 - q^{-1}e^{2/3}$  is marked. (In this plot, the values of $G'''(y)$ are indicated on the right-vertical axis, and the values of $G(y)$ are indicated on the left-vertical axis).
Thus $G''(y)$ is decreasing on $[1/q, y_0]$ and increasing on $[y_0,\theta]$.
Since $G''(\theta)=0$, it follows that $G''(y) <0$ on $[y_0,\theta)$.  We note that
\[ G''(1/q) =  \tfrac{\ln(q-1) (2 \theta-1)}{4 (\theta(1- \theta))^2} > 0.\]
Thus $G''(1/q) >0 > G''(y_0)$ together with the fact that $G''(y)$ is decreasing on $[1/q,y_0]$ implies that there is a unique $y_1$ in the interior of this interval such that $G''(y)$ has the same sign as $y_1-y$ on this interval. We have already shown that $G''(y) < 0$ on $[y_0,\theta]$. Thus we conclude
\[ \text{sign}(G''(y)) =  \text{sign}(y_1-y) \quad \text{ on } \; [1/q,\theta).\]
This implies $G'(y)$ is increasing on $[1/q,y_1]$ and decreasing on $[y_1, \theta]$.
Since $G'(\theta)=0$, we conclude that $G'(y) >0$ on $[y_1, \theta)$. We note that 
\[ G'(1/q)= \tfrac{-1}{\sqrt{\theta(1-\theta)}}  \int^{\sqrt{q-1}}_{1} \ln(t) (1 - \tfrac{1}{t^2})  dt  <0 .\]
Since $G'(y)$ is increasing on $[1/q,y_1]$  and $G'(1/q) <0 < G'(y_1)$, we conclude that there is a unique $y_2$ in the interior of the interval $[1/q,y_1]$ such that $G'(y)$ has the same sign as $y-y_2$ on this interval. Also $G'(y) >0$ on $[y_1, \theta]$. Thus we conclude:
\[ \text{sign}(G'(y)) =  \text{sign}(y-y_2) \quad \text{ on } \; [1/q,\theta).\]
This implies that  $G(y)$ is decreasing on $[1/q,y_2]$ and increasing on $[y_2, \theta]$.
Since $G(1/q) = G(\theta)=0$, we see that $G(y)$ is negative on $(1/q,y_2]$ as well as $[y_2, \theta)$. This finishes the proof of the assertion $G(y)<0$ on $(1/q,\theta)$, 
and hence of   \eqref{eq:MRRWmin1}.

Next, we prove the $\cup$-convexity of $\tilde \alpha_{MRRW}(x)$. We must show that the derivative $\tilde \alpha_{MRRW}'(x)$ is non-decreasing. Since the derivative is constant on $[0,(1 - 2/q)^2]$, the problem reduces to showing that   $\alpha_{MRRW}(x)$ is $\cup$-convex for $x \in [(1 - 2/q)^2,\theta]$. This follows from the next lemma:
\begin{lemma}  \label{MRRWconv} The first MRRW bound  $\alpha_{MRRW}(\delta)$ is 
$\cup$-convex if $q=2$. For $q>2$, it is $\cap$-convex on $[0,\delta_{MRRW}]$ and $\cup$-convex on $[\delta_{MRRW},\theta]$ where $\delta_{MRRW}$ satisfies:
\[ \tfrac{1}{2} - \tfrac{\sqrt{q-1}}{q} < \delta_{MRRW} < (1-\tfrac{2}{q})^2.\]
\end{lemma}
\bep Let $y = \xi(x)$. Let 
\[ \chi(y) = 1-2y+(2 \theta-1) \sqrt{\tfrac{y(1-y)}{\theta(1-\theta)}} \]
We calculate:
\beq \label{eq:A1}
\tfrac{y'^2}{2y'' y(1-y)} = \sqrt{\tfrac{x(1-x)}{\theta(1-\theta)}} =  \chi(y) 
\eeq
Since $\sqrt{x(1-x)/(\theta(1-\theta))}$ is non-negative, we also make the observation that  that $\chi(y)>0$   for all $y \in [0,\theta)$.
Since $\alpha_{MRRW}(x) =H_q(y)$, we get:
\[ \ln(q) \alpha_{MRRW}''(x) =  \tfrac{-y'^2}{y(1-y)} +  y''\, \ln \tfrac{(q-1)(1-y)}{y}.  \]
Since $\xi(\xi(x))=x$, we get $y' = \xi'(x) = 1/\xi'(y)$. Using this we get: 
\[ \alpha_{MRRW}''(x) (\xi'(y))^2 y(1-y) \ln(q)  =   -1  + \tfrac{2 y'' y(1-y)}{y'^2} \, \ln \sqrt{\tfrac{(q-1)(1-y)}{y}}.  \]
Using \eqref{eq:A1}, we obtain:
%\begin{multline*} \alpha_{MRRW}''(x) (\xi'(y))^2 y(1-y) \ln(q) = \\
%-1+ \tfrac{\ln  \sqrt{\tfrac{(1-y)(q-1)}{y}}}     {(2 \theta-1)   \sqrt{ \tfrac{y(1-y)}{\theta(1 - \theta)}} + 1 - 2y},
%\end{multline*}
\[ \label{eq: A2}  \alpha_{MRRW}''(x) (\xi'(y))^2 y(1-y) \ln(q) = 
-1+ \tfrac{\ln  \sqrt{\tfrac{(1-y)(q-1)}{y}}}     {\chi(y)} 
\]
Let $y \in (0,\theta)$.  We recall note   $\chi(y)> 0$ for $y \in (0,\theta)$. Thus  for  $\alpha_{MRRW}''(x)$ has the same sign as  
\[ G_2(y) := \ln (\sqrt{\tfrac{(1-y)(q-1)}{y}})  - \chi(y).  \]
We calculate:
\[G_2'(y) y(1-y) = \chi(y)(y-\tfrac{1}{2}).\] 
Therefore, sign$(G_2'(y)) = \text{sign}(y-1/2)$. In other words $G_2(y)$ is decreasing on $[0,1/2]$ and increasing on $[1/2,\theta)$.
We note $G_2(1/q) = \ln(q-1) - 2(1 - \tfrac{2}{q})$.
The function 
\[t \mapsto \ln(t-1) -  2(1 - 2/t),\]  evaluates to $0$ at $t=2$, and is an increasing function of $t$ for $t\geq 2$ (because  its derivative $(1-2/t)^2/(t-1)$ is positive). Thus $G_2(1/q)>0$ for $q>2$ and  $G_2(1/q)=0$ for $q=2$. Since  $G_2(0) = + \infty$   and $G_2(y)$ is decreasing on $[0,1/q]$, we conclude that $G_2(y) >0$ on $[0,1/q]$ if $q>2$. If $q=2$, then  $G_2(y) \geq 0$ on $[0,1/q]=[0, \theta]$. In particular, for $q=2$ the bound $\alpha_{MRRW}(x)$ is $\cup$-convex on $[0, \theta]$.\\

For $q>2$, we note that $G_2(1/2) = \tfrac{1}{2}( \ln(q-1) - \tfrac{q-2}{\sqrt{q-1}})$. The function $b(t) =  \tfrac{1}{2}( \ln(t-1) - \tfrac{t-2}{\sqrt{t-1}})$ satisfies  $b(3) =   \tfrac{1}{2}( \ln(2) - \tfrac{2}{\sqrt{2}})<0$ and $b'(t) = \tfrac{2 \sqrt{t-1} - t}{4 (t-1)^{3/2}} <0$ for $t \geq 3$. Thus $G_2(1/2) < 0$ for all $q>2$.
Since $G_2(1/q) >0$ and $G_2(1/2) <0$ and $G_2(y)$ is decreasing on $[1/q,1/2]$, we conclude that there is a $y_{MRRW} \in (1/q,1/2)$ such that sign$(G_2(y)) = \text{sign}(y_{MRRW} - y)$ for $y \in [0,1/2]$. Also $G_2(\theta) = 0$, $G_2(1/2)<0$ and $G_2(y)$ is increasing on $[1/2,\theta]$, which shows that $G_2(y) <0$ on $[1/2,\theta)$.
Thus sign$(G_2(y)) =\text{sign} (y_{MRRW} - y)$ for $y \in (0,\theta)$. Since  $\alpha_{MRRW}''(x)$ has the same  sign as $G_2(y)$ (where $y=\xi(x)$), we finally obtain sign$(\alpha_{MRRW}''(x))= \text{sign}(x -\delta_{MRWW})$ for $ x \in (0,\theta)$, where $\delta_{MRWW} = \xi(y_{MRRW})$  satisfies $\xi(1/2) < \delta_{MRWW} < \xi(1/q)$, or in other words: $\tfrac{1}{2} - \tfrac{\sqrt{q-1}}{q} < \delta_{MRRW} < (1-\tfrac{2}{q})^2$. This completes the proof of the lemma.
\eep

%\section*{Acknowledgment}
\bibliographystyle{IEEEtran}
\bibliographystyle{plain}
\bibliography{refs}
\nocite{}
\end{document}